\documentclass[reprint, 
superscriptaddress,
nofootinbib,
 amsmath,amssymb,
 aps, 
prb,
]{revtex4-2}
\usepackage[colorlinks=true, linkcolor=purple, citecolor=purple, urlcolor=blue]{hyperref}
\usepackage{xcolor}
\usepackage{graphicx}
\usepackage{dcolumn}
\usepackage{bm}
\usepackage{lipsum} 
\usepackage{float}      
\usepackage{physics}

\newcommand{\EqualContributionText}{These authors contributed equally to this work.}

\newcommand{\equalcontribmark}{$^{\ddagger}$}

\newcommand{\printequalcontrib}{%
  \begingroup
    \renewcommand{\thefootnote}{\fnsymbol{footnote}}%
    \footnotetext[3]{\EqualContributionText}%
  \endgroup
}

\begin{document}

\title{\textbf{Topological Surface Magnon-Polariton in an Insulating Canted Antiferromagnet} 
}%
\author{Weixin Li\equalcontribmark}
\affiliation{Fert Beijing Institute, MIIT Key Laboratory of Spintronics, School of Integrated Circuit Science and Engineering, Beihang University, Beijing 100191, China}

\author{Rundong Yuan\equalcontribmark}
\email{ry306@cam.ac.uk}
\affiliation{TCM Group, Cavendish Laboratory, University of Cambridge, J.\,J.\,Thomson Avenue, Cambridge CB3 0HE, United Kingdom}

\author{Fenglin Zhong}

\affiliation{CDT in Superconductivity, Cavendish Laboratory, University of Cambridge, J. J. Thomson Avenue, Cambridge CB3 0HE, United Kingdom}
\affiliation{Department of Materials Science and Metallurgy, University of Cambridge, Charles Babbage Road, Cambridge CB3 0FS, United Kingdom}

\author{Bo Peng}
\affiliation{TCM Group, Cavendish Laboratory, University of Cambridge, J.\,J.\,Thomson Avenue, Cambridge CB3 0HE, United Kingdom}

\author{Jean-Philippe Ansermet}
\affiliation{Shenzhen Institute for Quantum Science and Engineering, Southern University of Science and Technology, Shenzhen 518055, China}
\affiliation{Institute of Physics, École Polytechnique Fédérale de Lausanne (EPFL), 1015 Lausanne, Switzerland}

\author{Haiming Yu}
\email{haiming.yu@buaa.edu.cn}
\affiliation{Fert Beijing Institute, MIIT Key Laboratory of Spintronics, School of Integrated Circuit Science and Engineering, Beihang University, Beijing 100191, China}
\affiliation{International Quantum Academy, Shenzhen 518048, China}

\begin{abstract}
Excitation and control of antiferromagnetic magnon modes lie at the heart of coherent antiferromagnetic spintronics. Here, we propose a topological surface magnon-polariton as a new approach in the prototypical magnonic material hematite. We show that in an insulating canted antiferromagnet, where strong-coupled magnon-photon modes can be achieved using electrical on-chip layouts, a surface magnon-polariton mode exists in the gap of the bulk magnon-photon bands. The emergence of surface magnon-polariton mode is further attributed to the nontrivial topology of bulk magnon-photon bands. Magnon-photon coupling enhances the Berry curvature near the anticrossing points, leading to a topological bulk Chern band associated with the surface magnon-polaritons. Our work provides a general principle for the utilization of optomagnetic properties in antiferromagnets, with an illustration of its experimental feasibility and wide generality as manifested in hematite.

\end{abstract}

\maketitle
\printequalcontrib
\section{Introduction}
Magnons or spin waves, the collective excitation of magnetic moments, can carry and propagate spin precessional information in magnetic materials without any displacement of electrons, and thus can be utilized in spintronic devices with zero Ohmic loss~\cite{kruglyak2010magnonics, chumak2015magnon, barman2021roadmap, flebus2024, han2023coherent, ansermet2024spintronics}.
The application of magnons in electronic devices gives rise to the field of magnonics, which has been intensively investigated for ferro- or ferrimagnetic materials, such as metals permalloy~\cite{podbielski2006spin,chumak2009spin, yu2013omnidirectional}, CoFeB~\cite{yu2012high,rana2019electric, wang2022hybridized}, semimetal Co$_2$MnGa$_x$Ge$_{1-x}$~\cite{wang2025band}, semiconductor CrI$_3$~\cite{costa2020topological, cenker2021direct}, insulator yttrium iron garnet~\cite{serga2010yig, yu2014magnetic, wang2024broad}, and their heterostructures.
By contrast, given their (sub-)terahertz eigenfrequencies~\cite{han2023coherent, kampfrath2011coherent, li2020spin,  vaidya2020subterahertz, hortensius2021coherent}, antiferromagnetic magnons were mostly inaccessible to electronic control and detection. 
Significant progress in antiferromagnetic magnonics has been made only in recent years~\cite{ li2023ultrastrong, sun2024dipolar,  chen2025coherent, tang2025coherent, boventer2021room, wang2021spin, wang2023long, hamdi2023spin, el2023antiferromagnetic, chen2025observation, sheng2025control,el2025probing, su2025enhancement, lebrun2018tunable, li2020spin,vaidya2020subterahertz, wimmer2020observation,  hortensius2021coherent, boventer2021room, wang2021spin, bae2022exciton,chen2025observation, chen2025unconventional}.
In van der Waals antiferromagnets, gigahertz magnons have been observed~\cite{li2023ultrastrong,tang2025coherent,sun2024dipolar,chen2025coherent} due to their weak interlayer exchange interactions, despite their relatively high damping.
Recently, the rediscovery of the gigahertz antiferromagnetic magnon bands in an insulating canted antiferromagnet, hematite, arise a new area of antiferromagnetic magnonics, which possess advantages of high group velocity and insensitivity to external magnetic disturbances~\cite{boventer2021room, wang2021spin, wang2023long, hamdi2023spin, el2023antiferromagnetic}.
The interplay between strong antiferromagnetic exchange interactions, easy-plane anisotropy, and Dzyaloshinskii-Moriya interaction (DMI) results in magnon modes at the gigahertz range, and thus, making the electrical detection of magnons in hematite feasible~\cite{boventer2021room, wang2021spin, wang2023long, hamdi2023spin, el2023antiferromagnetic, chen2025observation, sheng2025control,el2025probing, su2025enhancement}.
These features, along with an outstandingly low damping coefficient~\cite{hamdi2023spin,su2025enhancement}, establish hematite as an ideal platform for antiferromagnetic magnonics. 
Hence, methods to excite magnon modes in hematite are desirable~\cite{han2023coherent, wang2023long, hamdi2023spin, el2023antiferromagnetic}, and here we propose an electrical approach targeting its photon-induced surface mode.

\begin{figure*}[t]
    \centering
    \includegraphics[width=\textwidth]{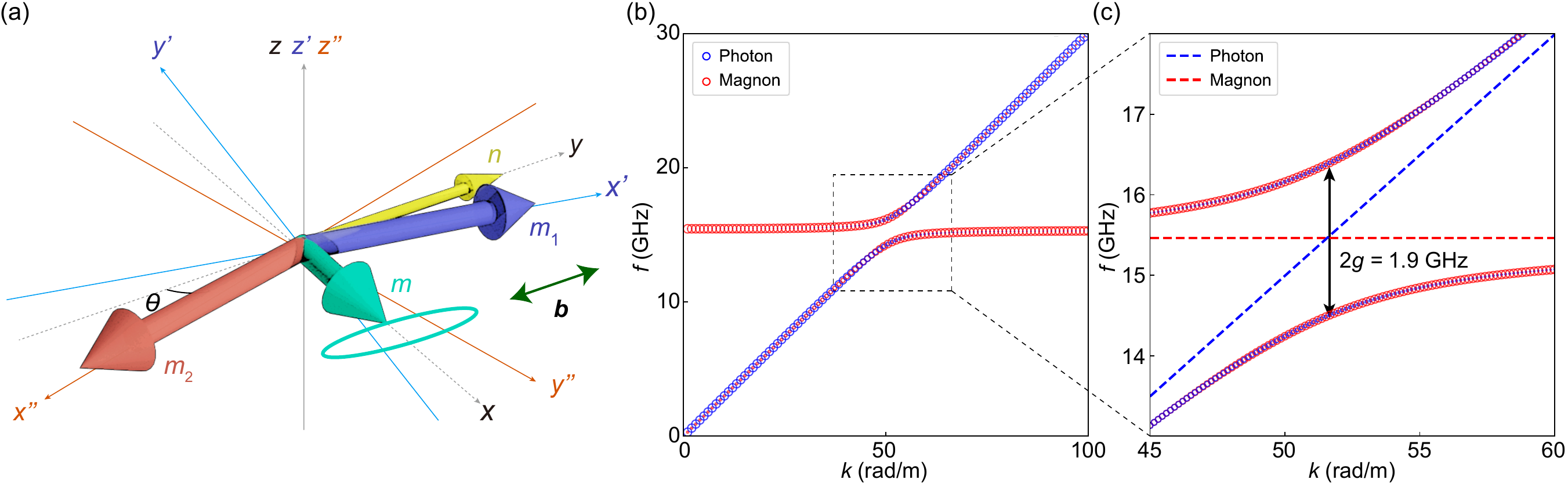}
    \caption{{{\bf Modelling bulk magnon-photon bands in the DMI-canted antiferromagnets.} (a) The DMI canted model of hematite with the frame of reference setting, sublattices $\boldsymbol{m}_1$, $\boldsymbol{m}_2$, net magnetic moment $\boldsymbol{m}=(\boldsymbol{m}_1+\boldsymbol{m}_2 )/2$, Néel vector $\boldsymbol{n}=(\boldsymbol{m}_1-\boldsymbol{m}_2)/2$, and an oscillating magnetic field $\boldsymbol{b}$ coupling to the in-plane polarized precession. (b) Component-projected dispersion relations of magnon-polaritons. The blue (red) circles denote the photon (magnon) component at the given wavevector and frequency, where the radius of the circle is proportional to the ratio of the corresponding component. (c) Zoomed-in view of the dispersion relations in the region marked by the dashed box in (b), where the blue (red) dashed lines denote the photon (magnon) dispersion without anticrossing. We use parameters that give a coupling strength $2g = 1.9~\text{GHz}$ to qualitatively fit the previous experimental observations~\cite{wang2025long}}.}
    \label{dispersion}
\end{figure*}

As the basis of this work, we emphasize that a strong-coupled magnon-photon mode with an extra-long propagation distance can be observed in a single-crystal hematite chip by using an electrical, cavity-free setup~\cite{wang2025long}.
Such a discovery enables the investigation of magnon-polaritons in hematite, particularly, the physical properties associated with their large and well-distinguished magnon-photon band gaps. Additionally, the on-chip magnon-photon coupling in a slab geometry provides the opportunity to investigate magnon modes located at the sample surface, whereas conventional cavity magnonics can only access the bulk resonance modes~\cite{boardman1982electromagnetic, bai2015spin, rameshti2022cavity, boventer2023antiferromagnetic,bialek2023cavity,rao2023meterscale}.
In easy-axis collinear antiferromagnets, it has been predicted that the surface magnon-polaritons can exist within the magnon-photon coupling gap in terahertz regime~\cite{macedo2019engineering}.
However, a general physical picture of such antiferromagnetic surface modes and its correspondence to bulk modes has not yet been established. Here, we propose that the nontrivial band topology gives rise to the emergence of surface magnon-polaritons in insulating DMI-canted antiferromagnets. Our claim is consistent with the notion of topological magnon-photon anticrossing in ferromagnets~\cite{shindou2013topological}, and with topological magnon-magnon coupling in artificial magnetic lattices~\cite{shindou2013topological,hu2022tunable,wang2023observation,pirmoradian2023topological}.

In this work, we theoretically investigate the topological surface magnon-polaritons in the insulating, DMI-canted, antiferromagnet hematite. We focus on two matters: first, whether the surface magnon-polaritons also exist in an easy-plane, DMI-canted antiferromagnetic ground state; second, whether the emergence of antiferromagnetic surface magnon-polaritons is related to the nontrivial topology of the magnon-photon bands. On top of our theoretical discussion, we also provide an experimental protocol for the design of microwave antenna to excite such surface mode. We use a semiclassical Landau-Lifshitz model combined with the Maxwell’s equations to account for the physics of antiferromagnetic magnons and photons in DMI-canted antiferromagnets. We show the manifestation of this model using hematite as a showcase system, but our model can be extended to any insulating antiferromagnet. 
The paper is organized as follows. In Sec.~\ref{model} we present a semiclassical model for bulk magnon-photon coupling. 
In Sec.~\ref{topo} we discuss the topology of the bulk magnon-photon bands. 
In Sec.~\ref{surface} we characterize the surface magnon–polariton modes. 
Finally, in Sec.~\ref{discuss} we summarize the results and propose experimental protocols.

\section{Semiclassical model for bulk magnon-photon coupling}\label{model}

We first present the semiclassical model for magnon-polaritons in insulating DMI-canted antiferromagnets. In a magnetic media, the modes of electromagnetic waves are governed by the magnetic susceptibility $\chi(\omega)$, which characterizes the motion of the net magnetic moment $\boldsymbol{m}(\omega)$ under an external excitation magnetic field $\boldsymbol{b}(\omega)$.
To account for the motion of $\boldsymbol{m}$, we consider a macrospin model with two sublattices $\boldsymbol{m}_1$ and $\boldsymbol{m}_2$, which are antiferromagnetically coupled, using the Landau-Lifshitz equation reads
\begin{widetext}
\begin{equation}
\begin{aligned}
\frac{d \boldsymbol{m}_{1}}{d t} & =-\gamma \mu_{0} \boldsymbol{m}_{1} \times\left[\boldsymbol{H}_{0}+\frac{\boldsymbol{b}}{\mu_0}-H_{\text{ex}} \boldsymbol{m}_{2}-H_{\text{A}}\left(\boldsymbol{m}_{1} \cdot \hat{z}\right) \hat{z}+H_{\text{a}}\left(\boldsymbol{m}_{1} \cdot \hat{y}\right) \hat{y}+H_{\text{DM}}\left(\boldsymbol{m}_{2} \times \hat{z}\right)\right], \\
\frac{d \boldsymbol{m}_{2}}{d t} & =-\gamma \mu_{0} \boldsymbol{m}_{2} \times\left[\boldsymbol{H}_{0}+\frac{\boldsymbol{b}}{\mu_0}-H_{\text{ex}} \boldsymbol{m}_{1}-H_{\text{A}}\left(\boldsymbol{m}_{2} \cdot \hat{z}\right) \hat{z}+H_{\text{a}}\left(\boldsymbol{m}_{2} \cdot \hat{y}\right) \hat{y}-H_{\text{DM}}\left(\boldsymbol{m}_{1} \times \hat{z}\right)\right],
\end{aligned}
\end{equation}
\end{widetext}
where $\boldsymbol{H}_{0}$ is the applied external magnetic field, $\boldsymbol{b} / \mu_{0}$ is the oscillating magnetic field of photon with $\mu_0$ the vacuum permeability,  $H_{\text {ex}}$ is the exchange coupling strength between two sublattices, $H_{\text{A(a)}} > 0 $ is the out-of-plane (in-plane) hard-axis (easy-plane) anisotropy coefficient, and $H_{\mathrm{DM}}$ is the DMI coefficient with the DMI vector oriented along $\boldsymbol{\hat{z}}$. 
The illustration of the frame of reference can be found in Fig.~\ref{dispersion}(a).
The exchange stiffness term proportional to $a_{ex}^2 k^2$ is ignored in this model based on the fact that magnon-photon anticrossing point is much smaller than the wavevector at first Brillouin zone (BZ) boundary, while the exchange stiffness $a_{ex}$ is at the same order of magnitude of lattice constant~\cite{wang2023long, hamdi2023spin,el2023antiferromagnetic}.
By introducing $\boldsymbol{m}=(\boldsymbol{m}_1+\boldsymbol{m}_2)/2$ and $\boldsymbol{n}=(\boldsymbol{m}_1-\boldsymbol{m}_2)/2$, we can switch the basis from $\boldsymbol{m}_1$, $\boldsymbol{m}_2$ to $\boldsymbol{m}$, $\boldsymbol{n}$.
Thus, we obtain the equations of motion that directly capture the physics of the net magnetic moment and the Néel vector, which is feasible for our analysis since only the net magnetic moment $\boldsymbol{m}$ is coupling to the photon. 
The details of the mathematical operations are further illustrated in Appendix~\ref{modelA}. 
Simultaneously, we have the Maxwell’s equations accounting for the physics of photons, which read~\cite{griffiths2023introduction}
\begin{equation}
    \begin{aligned}
        \frac{\partial \boldsymbol{b}}{\partial t} &= -\nabla \times \boldsymbol{e}, \\
        \frac{\partial \boldsymbol{e}}{\partial t} &= \frac{1}{\epsilon}\nabla \times \left(\frac{\boldsymbol{b}}{\mu_0}-M_{\textbf{s}}\boldsymbol{m}\right),
    \end{aligned}
\end{equation}
where $\boldsymbol{e}$ is the electric field of photon, $M_{\text{s}}$ is the magnetization of the canted moment, $\epsilon$ is the dielectric constant taken from experimental results~\cite{wang2025long}.
Here we only consider the displacement current contribution ($\epsilon \partial\boldsymbol{e}/\partial t$) of electric field in insulators to exclude the irrelevant surface plasmon-polariton solutions~\cite{griffiths2023introduction}.
After some algebra~(see Appendix~\ref{modelA}), the effective Schrödinger equation from Landau-Lifshitz and Maxwell’s (LLM) equations reads
\begin{equation}\label{eigenEq}
\mathcal{H}_{\text{eff}}\boldsymbol{x_k}^{(i)} = \omega_i \boldsymbol{x_k}^{(i)},
\end{equation}
where $mathcal{H}_{\text{eff}}$, is the effective Hamiltonian of bulk magnon-photon modes (see Appendix~\ref{modelA}), $\omega_i$ is the eigenfrequency of the band with index $i$, and $\boldsymbol{x_k}^{(i)}\equiv\left[m_{\boldsymbol{k},y}^{(i)},m_{\boldsymbol{k},z}^{(i)},n_{\boldsymbol{k},x}^{(i)}, \boldsymbol{b_k}^{(i)}, \boldsymbol{e_k}^{(i)}\right]^T$ denotes the eigenvector of the band with index $i$.
Here, we only consider the solution with positive frequency, and we find all frequencies are real. 
he upper and lower bulk magnon-photon bands [$f_{1,2}=\omega_{1,2}/(2\pi)$] are shown in Fig.~\ref{dispersion}(b)-(c), where the parameters we take are present in Appendix~\ref{modelA}.
We choose the value of $M_{\text{s}}$ to qualitatively fit the experimental observations~\cite{wang2025long}.
The eigenvectors are normalized by~\cite{okamoto2020berry} $\norm{\boldsymbol{x_k}^{(i)}}^2=\mathcal{E}_i$, where $\mathcal{E}_i = \hbar \omega_i$ denotes the energy of the magnon-photon mode with band index $i$.
In Fig.~\ref{dispersion}(b), we show the component-projected magnon-polariton band structure, where the red (blue) circles indicate the relative contributions of the magnon (photon) components in the wavefunction by $r=\mathcal{E}_i^{\text{mag(ph)}}/\mathcal{E}_i$ with $\mathcal{E}_i^{\text{mag(ph)}}$ the energy from magnon (photon) components.
The conversion between magnon and photon components is analogous to the behavior of the hybridized electronic orbitals near the band-inversion-induced anticrossing gap in topological insulators~\cite{hsieh2012topological, qian2014quantum, chang2023colloquium}, indicating the possible emergence of nontrivial topology in our system. 
One can also find a strong magnon-photon coupling captured by its minimal band gap $2g=1.9~\text{GHz}$, as shown in the zoom-in plot of the dispersion relation in Fig.~\ref{dispersion}(c).
Furthermore, in our model, the eigen precession mode of $\boldsymbol{m}$ possesses an in-plane polarization in agreement with previous works, as drawn in Fig.~\ref{dispersion}(a), which further induces an anisotropic coupling strength as illustrated in Appendix~\ref{anisotropic_A}.

Our semiclassical model characterizes the strong magnon-photon coupling in hematite~\cite{boventer2021room, wang2025long}, even with a small saturated magnetization $M_{\text{s}}$ (see Appendix~\ref{modelA}). 
If one takes the parameters of other materials with appropriate approximations accordingly, the methodology here can be instantly applied to account for magnon-photon coupling in other insulating antiferromagnetic materials. 
However, given that each antiferromagnet possesses different eigen precession modes, the involved components in the eigenvector $\boldsymbol{x_k}^{(i)}$ and the frame-of-reference transformation should also be adjusted accordingly when applying such approach to other materials.

\section{Nontrivial topology of bulk magnon-photon bands}\label{topo}

\begin{figure}
    \includegraphics[width=0.48\textwidth]{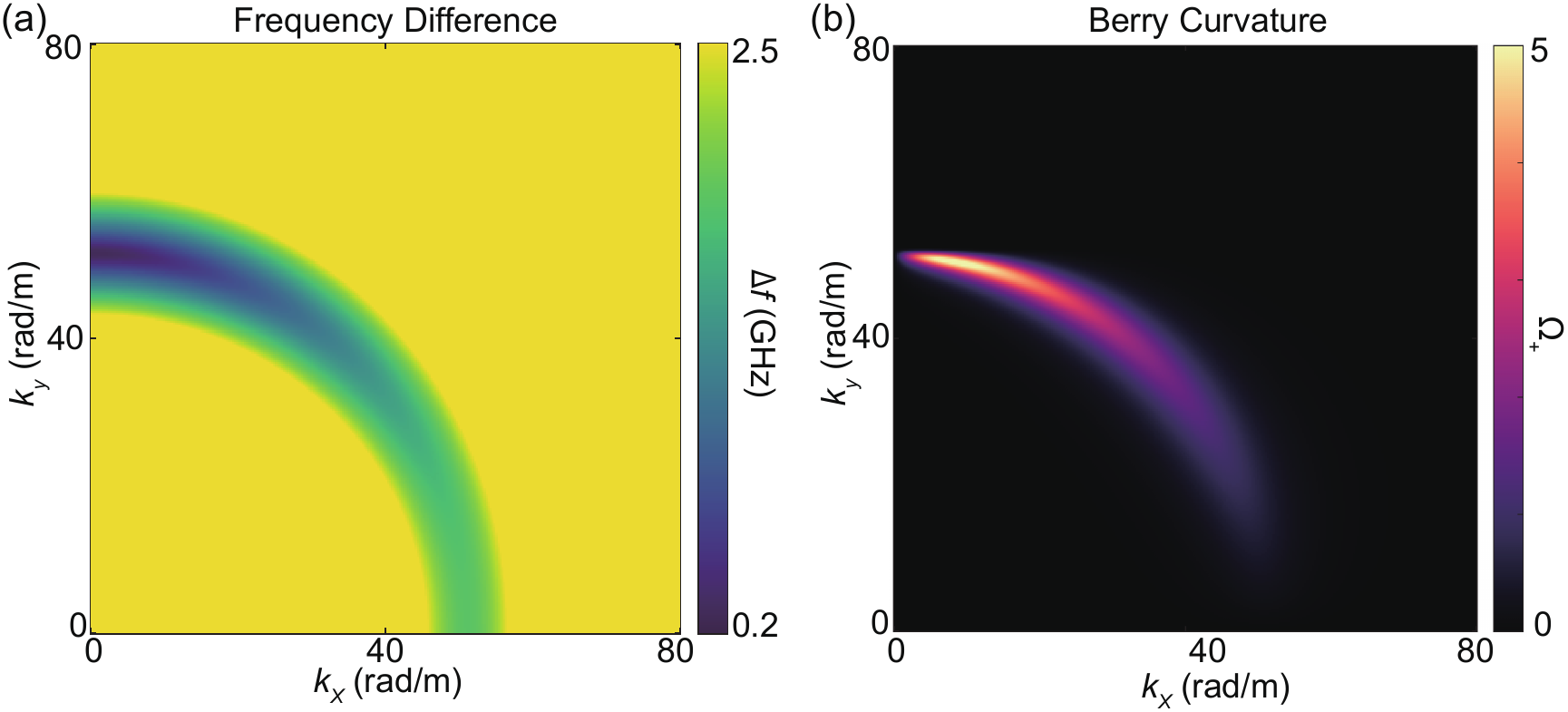}
    \caption{{\bf Distribution of frequency difference $(f_1 - f_2)$ and Berry curvature in reciprocal space.} (a) Distribution of frequency difference in reciprocal space with a color scale. The minima occur around the anticrossing points. (b) Distribution of the Berry curvature in reciprocal space. The Berry curvature is enhanced around the minima of bulk band gaps.}
    \label{berry}
\end{figure}
We then discuss the topological properties of the bulk magnon-photon bands. 
First, we find that the lower bulk magnon-photon band possesses trivial topology. 
The lower bulk magnon-photon band touches the vacuum photon mode at the $\Gamma$ point (with zero frequency), which indicates that the lower bulk magnon-photon band is topologically equivalent to the vacuum, or in other words, the lower bulk magnon-photon band carries a zero Chern number $C=0$. 
We then investigate the topology of the upper bulk magnon-photon band which does not touch with any other bands in the energy range of our interest.
This band could also touch other higher-energy bands of magnons in hematite~\cite{hoyer2025altermagnetic}, however here we assume the band touching is faraway that do not affect the topological property around $15~\text{GHz}$.
We calculate the Berry curvature which is defined as follows~\cite{berry1989quantum,owerre2017noncollinear,owerre2017topological,okamoto2020berry,Bostrom2023, yuan2025quantum}
\begin{equation}\label{berryEq}
    \Omega(\boldsymbol{k})=-2\operatorname{Im} \frac{\bra{\boldsymbol{x_k}^{(1)}} \partial_{k_{x}} \mathcal{H}_{\text{eff}}\ket{\boldsymbol{x_k}^{(2)}}\bra{\boldsymbol{x_k}^{(2)}} \partial_{k_{y}} \mathcal{H}_{\text{eff}}\ket{\boldsymbol{x_k}^{(1)}}}{\hbar\left(\omega_{1}-\omega_{2}\right)^{2}}.
\end{equation}
Among the eigenvectors, we find only the coupling two branches possess overlap, which justify the Eq.~\ref{berryEq}.
The format of Berry curvature is chosen to eliminate the gauge problem~\cite{owerre2017noncollinear,owerre2017topological,okamoto2020berry,Bostrom2023, yuan2025quantum}.
Given that our focus is on the topological surface magnon-polariton mode at the surface perpendicular to $\hat{\boldsymbol{z}}$, in Eq.~\ref{berry}, and the following analysis, we set $k_z=0$ in the Hamiltonian. 
In Fig.~\ref{berry}(a), we show the distribution of frequency difference~$\Delta f = f_2 - f_1$ between upper and lower bulk magnon-photon between upper and lower bulk magnon-photon band, and in Fig.~\ref{berry}(b) we show the Berry curvature of the upper band.
As shown in Fig.~\ref{berry}(a), the frequency difference varies as a function of the wavevector direction in consistence with our discussion on anisotropic coupling (see Appendix~\ref{anisotropic_A}), reaching minima at the anticrossing points. 
In Fig.~\ref{berry}(b), we can find that for the upper bulk magnon-photon band, the magnitude of Berry curvature $\Omega_+$ is significantly enhanced where the band gap is minimal. This can be understood by the fact that at the anticrossing point with the smallest band gap, the eigenvector switching from magnon-dominated to photon-dominated most rapidly, as demonstrated in Fig.~\ref{dispersion}(b)-(c).
The rapid mode shifting provides a large differentiation of the eigenvector $\left[\partial_{\boldsymbol{k}} \ket{\boldsymbol{x}^{(i)}}\right]$, and further contribute to the local Berry curvature. 
Also, the shifting of eigenvector is potential to provide nontrivial winding in the reciprocal space, as characterized by the Chern number.

We further calculate the Chern number of the upper bulk magnon-photon band by an integration of Berry curvature in the first BZ~\cite{shindou2013topological, okamoto2020berry}: $C_+=\int_{\text{BZ}} \Omega(\boldsymbol{k}) d \boldsymbol{k}/(2\pi)=1.$
We assume the boundary of the first BZ is far away from the dispersion we investigated here, and we only integrate around $\boldsymbol{k}$ points with nonzero Berry curvature~\cite{note1}.
The nonzero Chern number confirms that the upper bulk magnon-photon band is topologically nontrivial. 
Anticrossing between the bulk magnon and photon modes acts as a source of the Berry curvature, and further gives rise to the nontrivial topology of the upper band, which is the main claim of the nontrivial band topology in this section.

Given the nontrivial upper magnon-photon bulk band alongside with the topologically trivial lower band, we could predict that in the gap of bulk magnon-photon band there should exist an emergent mode. 
In real space, the vacuum is topologically trivial, and the bulk of hematite possesses a topologically nontrivial magnon-photon mode. 
Thus, at the interface of hematite and vacuum, the aforementioned emergent mode should appear, which can be concluded as a surface mode given our slab geometry. 
In our system, such discussions on bulk-boundary correspondence can be interpreted as the sign to the existence of topological surface magnon-polariton, as further discussed in the following text.

\section{Surface magnon-polariton mode}\label{surface}
\begin{figure}
    \includegraphics[width=0.48\textwidth]{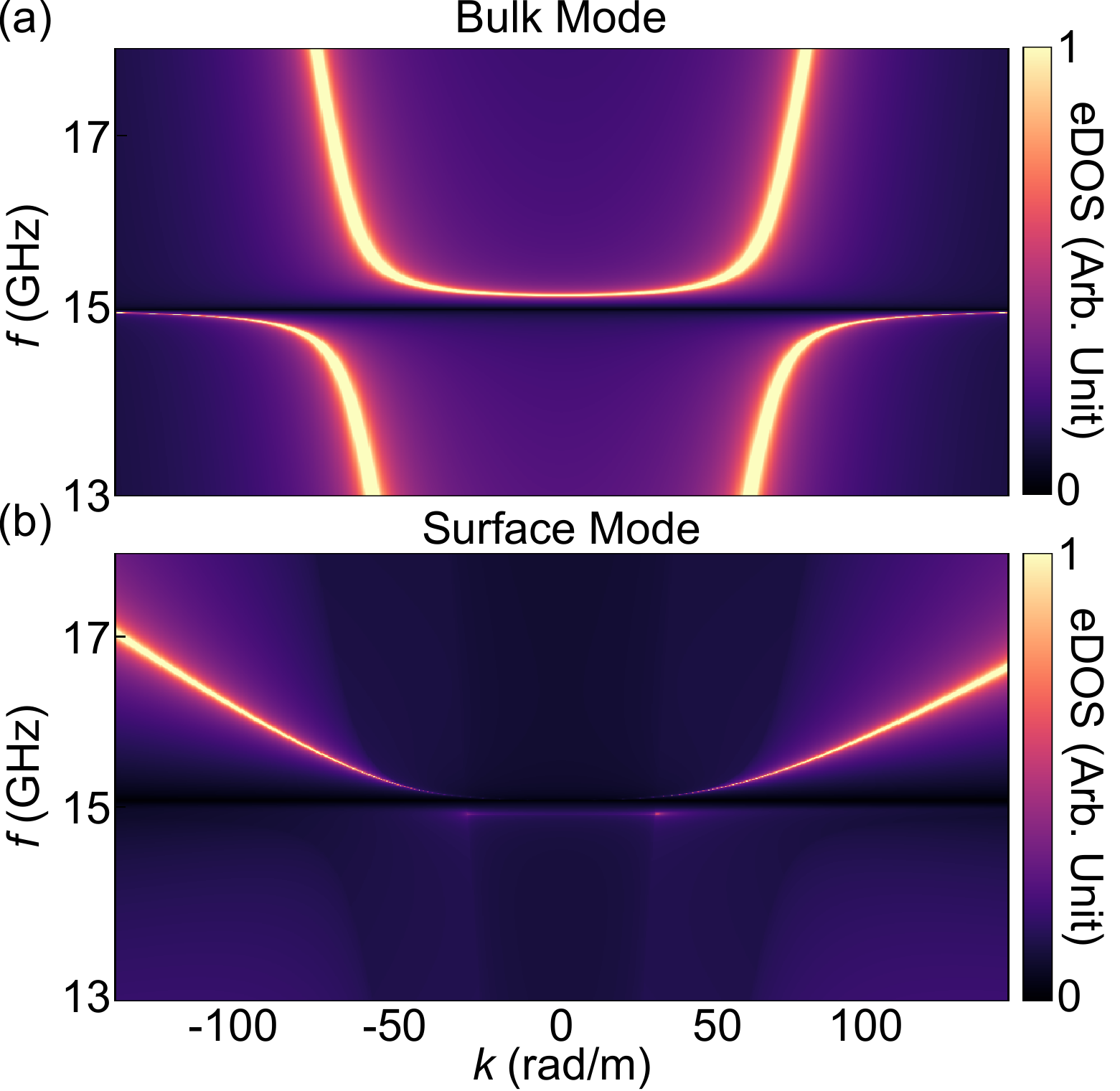}
    \caption{{\bf Reciprocal bulk magnon-polariton and nonreciprocal surface magnon-polariton.} (a) The effective DOS (eDOS) of magnon-polariton with a bulk Ansatz (see Appendix~\ref{surface_A}) which reproduce the magnon-photon band and its coupling strength in a color scale. The yellow color indicates a mode locating at the given wavevector and the frequency with a similar data process procedure in Ref.~\cite{macedo2019engineering}. (b) The eDOS of surface magnon-polariton with a finite evanescent wavevector $q$ and an $x$-direction propagation at the upper surface in a color scale with the same meaning of (a). The surface magnon-polariton appears within the bulk band gap and possesses nonreciprocity which can be tuned by the external magnetic field.}
    \label{surface_fig}
\end{figure}
In this section, we exhibit the surface magnon-polariton mode from our semiclassical approach.
Given our prediction, we conduct calculations for the surface magnon-polariton mode from our semiclassical approach. Our methodology follows Ref.~\cite{macedo2019engineering}, where they also find surface magnon-polariton modes in collinear antiferromagnets.
We provide the details of calculations in Appendix~\ref{surface_A}, where especially we provide the methodology to deduce the effective density of states (eDOS) of surface and bulk magnon-polaritons.
The maxima of eDOS indicate a coherent mode occurs at the given wavevector and frequency.
Here, inside the magnet, we consider the physics of the magnon-coupled electromagnetic waves at the surface, with an evanescent wavevector $q$ perpendicular to the slab surface providing a factor of amplitude $e^{qz}$.
The evanescent wavevector should obey the magnetostatic condition that $\nabla\cdot \boldsymbol{b} =0$. 
Meanwhile, we have the continuous condition of magnetic field at the surface $\partial_{\parallel}\psi_{\text{in}}=\partial_{\parallel}\psi_{\text{out}}$, where  $\psi_{\text{in(out)}}$ is the magnetic scalar potential inside(outside) the sample.
As further illustrated in Appendix~\ref{surface_A}, the surface mode dispersion follows the equation
\begin{equation}\label{SurfDis}
    \beta+\frac{q \mu_{33}-ik\mu_{23}}{\mu_{22}\mu_{33}-\mu_{23}\mu_{32}}=0,
\end{equation}
where $\beta=\sqrt{k^2-\omega^2/c^2}$, with $c$ the light velocity in our dielectric media, and $\mu_{ij}$ denotes the matrix components of dynamic magnetic permeability $\mu(\omega)$, which can be derived from Eq.~\ref{LL}, again, whose details are exhibited in Appendix~\ref{surface_A}.
Here, $\beta$ is real since the frequency of surface magnon-polaritons is predicted to appear in the gap, and thus to be lower than the light frequencies.
In our calculations, we take the same parameter of other sections, and we choose the propagation direction to be $x$ with an $1.9~\text{GHz}$ coupling gap.
If we take $\text{Re}(q)=0$, we can also deduce the bulk dispersion of magnon-photon bands in hematite, whose details are also exhibited in Appendix~\ref{surface_A}.
As illustrated in Fig.~\ref{surface_fig}(a), the well-reproduced bulk dispersion appears at the same frequency of Fig.~\ref{dispersion}(b)-(c), which verifies the legitimacy of our methodology, and further confirms the existence of band gap of magnon-polaritons in hematite.

Following Eq.~\ref{SurfDis}, with a finite evanescent wavevector $q$, we obtain the dispersion relation of surface magnon-polaritons in hematite, which are shown by the maxima of eDOS in Fig.~\ref{surface_fig}(b).
In agreement with the expectation for the topological bulk band as discussed above, the surface mode appears in the band gap between the topologically trivial and nontrivial bands. 
In electronic systems, the topological surface (edge) modes tend to connect the valence and conduction bands of electrons. 
In the magnetic case considered here, the physical interpretation of this surface mode is different. 
The upper band of magnon-polariton gradually approaches the light-velocity when $k$ approaches 100 rad/m, which is the largest group velocity of any (quasi-)particle, making it impossible for surface magnon-polariton to connect the ``valence'' (lower) bulk band and the ``conduction'' (upper) bulk band. 
The group velocity of the surface magnon-polaritons in hematite ($\sim10000~\text{km/s}$) is still much higher than that of pure magnons ($\sim20~\text{km/s}$)~\cite{wang2023long,hamdi2023spin, el2023antiferromagnetic}.
In the polaritonic regime, the magnetic mode appears as if it ``borrowed'' some property of the other mode (photons) with which it is in resonance, which has been found in hematite~\cite{wang2025long}.
Likewise, the surface polariton exhibits a long propagation distance, making detection of such modes possible. 

One can also observe a nonreciprocal dispersion of the surface mode in Fig.~\ref{surface_fig}(b) and is further illustrated in Appendix~\ref{NR_A}.
The nonreciprocal behavior is similar to the nonreciprocity found in Ref.~\cite{macedo2019engineering}, which originates from the external magnetic field canting the sublattice magnetizations. 
The surface magnon-polariton investigated here is already canted by the DMI in the system and thus induce nonreciprocity inherently. The nonreciprocity can be enhanced by the external field, as shown in Appendix~\ref{NR_A}.
The nonreciprocity depends on the evanescent wavevector $q$, which indicates a different propagation preference on either upper or lower surface. This chirality of surface state is analogous to the chirality of edge states in quantum (spin) Hall systems~\cite{laughlin1981quantized, kane2005z}, in conformity with the topological origin of the surface magnon-polariton.

\section{Discussion and Conclusions}\label{discuss}
In this section, we first give experimental protocols on the excitation and detection of topological surface magnon-polaritons.
The excitation of magnons in materials requires a match of wavevector which is determined by the shape factor of the microwave antenna~\cite{chumak2009spin, yu2013omnidirectional, yu2012high, rana2019electric, wang2022hybridized, wang2025band}.
Here, we propose in addition that the excitation of the topological surface magnon-polariton requires the match of both in-plane wavevector $\boldsymbol{k}$ and the evanescent wavevector $q$. 
In Appendix~\ref{match_A}, we show how the evanescent wavevector $q$ evolves as a function of wavevector $\boldsymbol{k}$.
The microwave waveguides with a prism-like total internal reflection can be a candidate for the excitation of topological surface magnon-polariton~\cite{macedo2019engineering}, where we can design simultaneously an in-plane propagating wavevector and an out-of-plane evanescent wavevector. 
Then, we further propose that the detection of such modes requires a time-resolved technology of microwave. 
A microsecond level resolution is required for a millimeter propagation length to distinguish such mode from bulk modes, after eliminating the signals from cross talks. 
Applications can be investigated by using the mode to give extra long-range coherent magnon propagation and/or interfere with other modes to provide different magnon precession~\cite{sheng2025control, lee2008conceptual, schneider2008realization, chen2021reconfigurable}, where one can refer to the polarization of the surface magnon-polaritons shown in Appendix~\ref{polarization_A}.
Although the results are deduced in the uses of the parameters from hematite, the physics captured in this work is not sensitive to the exact values of parameters but general to any system with the same low-loss and canting configuration of magnetism. 
Further investigations can be carried out for mode with a quantum origin for either magnon or photon beyond semiclassical LLM methodology, which can be investigated using second-quantized approach of magnon-photon Hamiltonian and quantum transport theories~\cite{Shen2020magnon,yuan2022quantum}.

To sum up, we discuss an emergent magnon-photon mode, topological surface magnon polariton, in an insulating canted antiferromagnet using the parameter of hematite.
Our semiclassical model reproduces the strong coupling between magnon and photon occurs in the bulk of such systems. 
The anticrossing behavior of bulk magnon-photon band is further confirmed as the origin of the nontrivial Berry curvature of the system. 
A nonzero Chern number $C=1$ is found for the upper bulk band and hosts the existence of topological surface mode in the system. 
Adopting our semiclassical model with an evanescent Ansatz across the surface, we find the predicted surface mode appears within the bulk band gap with a nonreciprocal feature.
Our work extends the methodology to investigate and utilize magnons in antiferromagnetic magnonic devices and provides a further function for the strong magnon-photon coupling in antiferromagnets.

\section*{Acknowledgments}
The authors thank Robert-Jan Slager, Zhejunyu Jin for fruitful discussions, and Hanchen Wang for his help on figure drawing. We acknowledge the support by the National Key Research and Development Program of China, Grant No. 2022YFA1402801; NSF China under Grants No. 12474104 and No. 52450018. R.Y. acknowledges the support by CSC Scholarship No. 202408060249 and Cambridge Commonwealth, European, and International Trust. F.Z. acknowledges the support by EPSRC Center for Doctoral Training in Superconductivity (Grant No. EP/Y035453/1). B.P. acknowledges support from Magdalene College Cambridge for a Nevile Research Fellowship.

\section*{Data Availability}
The data are available from the authors upon reasonable request.

\appendix

\section{Details for semiclassical Model of magnon-photon coupling}\label{modelA}
By introducing $\boldsymbol{m}=(\boldsymbol{m}_1+\boldsymbol{m}_2)/2$ and $\boldsymbol{n}=(\boldsymbol{m}_1-\boldsymbol{m}_2)/2$, we can switch the basis from $\boldsymbol{m}_1$, $\boldsymbol{m}_2$ to $\boldsymbol{m}$, $\boldsymbol{n}$, and deduce equations of motion that directly capture the physics of net magnetic moment ($\boldsymbol{m}$) and the Neel vector ($\boldsymbol{n}$), which is feasible for our analysis focusing on the motions of $\boldsymbol{m}$, since  only $\boldsymbol{m}$ is the real magnetic moment coupled with the photon.
After the basis transformation we have
\begin{widetext}
\begin{equation}\label{LL}
\begin{aligned}
\frac{d\boldsymbol{m}}{dt} = -\gamma \mu_0 \Bigg[ &\boldsymbol{m} \times \boldsymbol{H}_0 + \boldsymbol{m} \times\frac{\boldsymbol{b}}{\mu_0} +H_{\text{DM}} \Big( (\boldsymbol{n} \cdot \hat{z}) \boldsymbol{m} - (\boldsymbol{m} \cdot \hat{z}) \boldsymbol{n} \Big) - H_\text{A} \Big( (\boldsymbol{m} \cdot \hat{z}) (\boldsymbol{m} \times \hat{z}) + (\boldsymbol{n} \cdot \hat{z}) (\boldsymbol{n} \times \hat{z}) \Big) \\
& + H_{\text{a}} \Big( (\boldsymbol{m} \cdot \hat{y}) (\boldsymbol{m} \times \hat{y}) + (\boldsymbol{n} \cdot \hat{y}) (\boldsymbol{n} \times \hat{y}) \Big) \Bigg], \\
\frac{d\boldsymbol{n}}{dt} = -\gamma \mu_0 \Bigg[& \boldsymbol{n} \times \boldsymbol{H}_0 + \boldsymbol{n} \times\frac{\boldsymbol{b}}{\mu_0} + 2H_{\mathrm{ex}} (\boldsymbol{m} \times \boldsymbol{n}) + H_{\mathrm{DM}} \Big( (\boldsymbol{m} \cdot \hat{z}) \boldsymbol{m} - (\boldsymbol{n} \cdot \hat{z}) \boldsymbol{n} \Big)  \\&- H_{\mathrm{A}} \Big( (\boldsymbol{m} \cdot \hat{z}) (\boldsymbol{n} \times \hat{z}) + (\boldsymbol{n} \cdot \hat{z}) (\boldsymbol{m} \times \hat{z}) \Big) 
 + H_{\mathrm{a}} \Big( (\boldsymbol{m} \cdot \hat{y}) (\boldsymbol{n} \times \hat{y}) + (\boldsymbol{n} \cdot \hat{y}) (\boldsymbol{m} \times \hat{y}) \Big) \Bigg].
\end{aligned}
\end{equation}
\end{widetext}
Here, we also rewrite the Maxwell equations as
\begin{equation}
\begin{aligned}
\frac{\partial \boldsymbol{b}}{\partial t}&=-i  K \boldsymbol{e}, \\
    \frac{\partial \boldsymbol{e}}{\partial t}&=i \frac{K}{\epsilon}(\frac{ \boldsymbol{b}}{\mu_0}-M_{\text{s}}\boldsymbol{m}).
\end{aligned}
\end{equation}
 where matrix $K$ is defined as:
\begin{equation}
        K = \begin{pmatrix}
            0 & -k_z & k_y \\
            k_z & 0 & -k_x \\
            -k_y & k_x & 0
        \end{pmatrix}
\end{equation}

Substituting this frame-of-reference transformation into Landau-Lifshitz equations, also combining with the Maxwell's equations, we derive the Landau-Lifshitz-Maxwell equations of motion in matrix form that reads
\begin{equation}
    i \frac{\partial}{\partial t} \boldsymbol{x}_{\boldsymbol{k}}=\mathcal{H}_{\mathrm{eff}} \boldsymbol{x}_{\boldsymbol{k}}.
\end{equation}
From the LLM equations, the linearized dynamics reveal that the time derivatives of the components $m_{\boldsymbol{k},x}$, $n_{\boldsymbol{k},y}$ and $n_{\boldsymbol{k},z}$ are zero when taking into consideration only the low frequency mode, which is our main focus.
$\mathcal{H}_{\mathrm{eff}}$
follows:
\begin{equation}\label{hamiltonian}
\mathcal{H}_{\text{eff}}=\left(\begin{array}{ccc}
\mathcal{H}_{3 \times 3}^{\text {mag }} & -\theta \gamma\sigma_{2}^{\prime} & O_{3 \times 3} \\
O_{3 \times 3} & O_{3 \times 3} & K_{3 \times 3} \\
\theta \frac{K^{\prime}}{\epsilon_{0}} & -\frac{K_{3 \times 3}}{\epsilon_{0} \mu_{0}} & O_{3 \times 3}
\end{array}\right),
\end{equation}
where we have $O$ denoting the zero matrices, and 

\begin{subequations}\label{LLMDef}
    \begin{align}
    \mathcal{H}^{\text{mag}}_{3\times 3} &= -\gamma \mu_0
        \begin{pmatrix}
            0 & iH_1 & 0 \\
            iH_2 & 0 & iH_3 \\
            0 & iH_4 & 0
        \end{pmatrix},
    \\
        K^{\prime} &= \begin{pmatrix}
            -k_z & k_y & 0\\
             0 & -k_x & 0\\
             k_x & 0 & 0
        \end{pmatrix},
        \end{align}
    \begin{align}
        \sigma_2^{\prime} &=
        \begin{pmatrix}
            0 & 0 & -i \\
            0 & i & 0  \\
            0 & 0 & 0
        \end{pmatrix},
    \\
        \theta &= \frac{H_0 + H_{\text{DM}}}{2 H_{\text{ex}}}.
    \end{align}
\end{subequations}
In Eqs.~\ref{hamiltonian}-\ref{LLMDef}, $\theta$ is the canting angle between the N\'eel vector and the equilibrium value of $\boldsymbol{m}_1$, which can be found by minimizing the energy of static model, see also~\cite{wang2023long}.
We have $\theta =\arcsin (H_{\text{DM}}+H_0)/(2H_{\text{ex}})\ll 1$ since the $H_0$ and $H_\text{DM}$ is much smaller than $H_\text{ex}$ in hematite.
However, if one discusses such model in other materials, this approximation should be reexamined.
For the definitions in Eq.~\ref{LLMDef}, we have
\begin{subequations}
    \begin{align}
H_1&= \frac{H_0 + H_{\text{DM}}}{2 H_{\text{ex}}} H_{\text{A}} + H_0 - H_{\text{DM}}, \\
H_2 &= \frac{H_0 (H_{\text{a}} - 2 H_{\text{ex}}) + H_{\text{a}} H_{\text{DM}}}{2 H_{\text{ex}}},\\
H_3 &= H_{\text{a}} ,\\
H_4 &= \frac{(H_0 + H_{\text{DM}}) H_{\text{DM}} - 2 H_{\text{ex}} (H_{\text{a}} + H_{\text{A}} + 2 H_{\text{ex}}) }{2 H_{\text{ex}}}.
\end{align}
\end{subequations}
In our calculations, we set $H_0= 60~\text{mT}$ , $H_{\text{ex}} = 1040~\text{T}$, $H_{\text{DM}} = 2.7~\text{T}$, $H_{\text{a}} = 0.067~\text{mT}$, $H_{\text{A}} = 1~\text{mT}$ and $M_\text{s}$ = 1000~\text{A/m} to represent the magnetic properties of hematite~\cite{boventer2021room,wang2021spin, wang2023long}.

\section{Anisotropic Magnon-Photon Coupling Strength}\label{anisotropic_A}
\begin{figure*}
    \includegraphics[width=0.86\textwidth]{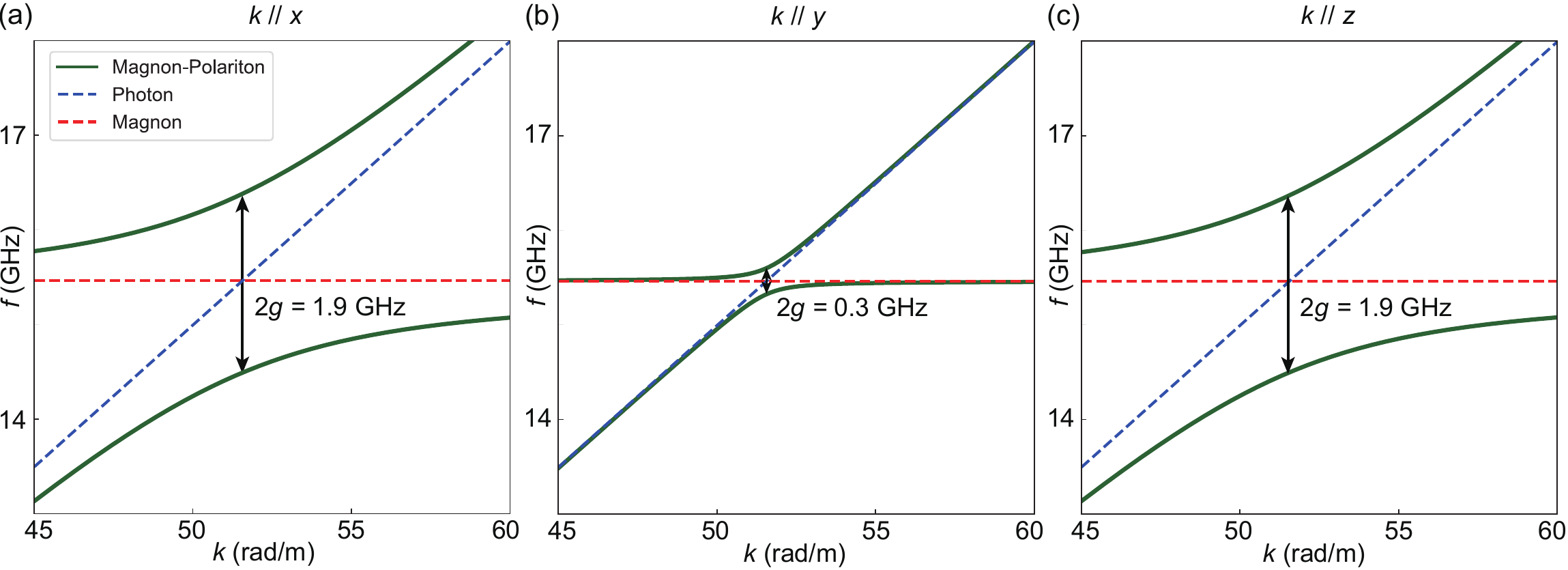}
    \caption{{\bf Anisotropic coupling strength of the bulk magnon-photon bands.} Dispersion relations showing the coupling strength along different wavevector directions: (a) for $\boldsymbol{k}//\hat{\boldsymbol{x}}$, (b) for $\boldsymbol{k}//\hat{\boldsymbol{y}}$ and (c) for $\boldsymbol{k}//\hat{\boldsymbol{z}}$. The minimal band gap occurs along the $k_y$ direction.}
    \label{anisotropic_fig}
\end{figure*}
In this appendix, we give a detailed discussion on the origin of the anisotropic phenomena that occur in the system. 
As shown in Fig.~\ref{anisotropic_fig}, the band gap exhibits a pronounced anisotropy, being minimal along the $k_y$ direction. Our calculations reveal that the gap sizes along the $k_x$ and $k_z$ directions are approximately six times larger than that along the $k_y$ direction, with a ratio of roughly 6:1. This anisotropy directly stems from the anisotropy of the magnonic modes in the canted antiferromagnets, specifically from the elliptical precession of the net magnetic moment $\boldsymbol{m}$.

For a DMI-canted antiferromagnetic ground state, the GHz excitation mode of $\boldsymbol{m}$ is elliptical precession, while for $n$ it is linear oscillations~\cite{wang2021spin,boventer2021room}. 
Given the magnon-photon coupling is mediated by the Zeeman coupling term $\boldsymbol{m}\cdot\boldsymbol{b}$, the energy scale varies significantly when the oscillating magnetic field of photon, $\boldsymbol{b}$, possesses different magnitude on $y$ or $z$ direction, i.e. has different polarizations. 
When the photon wavevector is along $y$ direction, the coupling interaction is minimized, since the photon magnetic field aligns along $x$ and/or $z$ direction, since the oscillating $\boldsymbol{m}$ is majorly along $y$. 
Similarly, the coupling strength when $\boldsymbol{k}$ is along $x$ and $z$ is maximized, since the photon magnetic field has $y$ component.

\section{Wavevector-dependent polarization of bulk and surface magnon-polaritons}\label{polarization_A}
\begin{figure}[h]
    \includegraphics[width=0.48\textwidth]{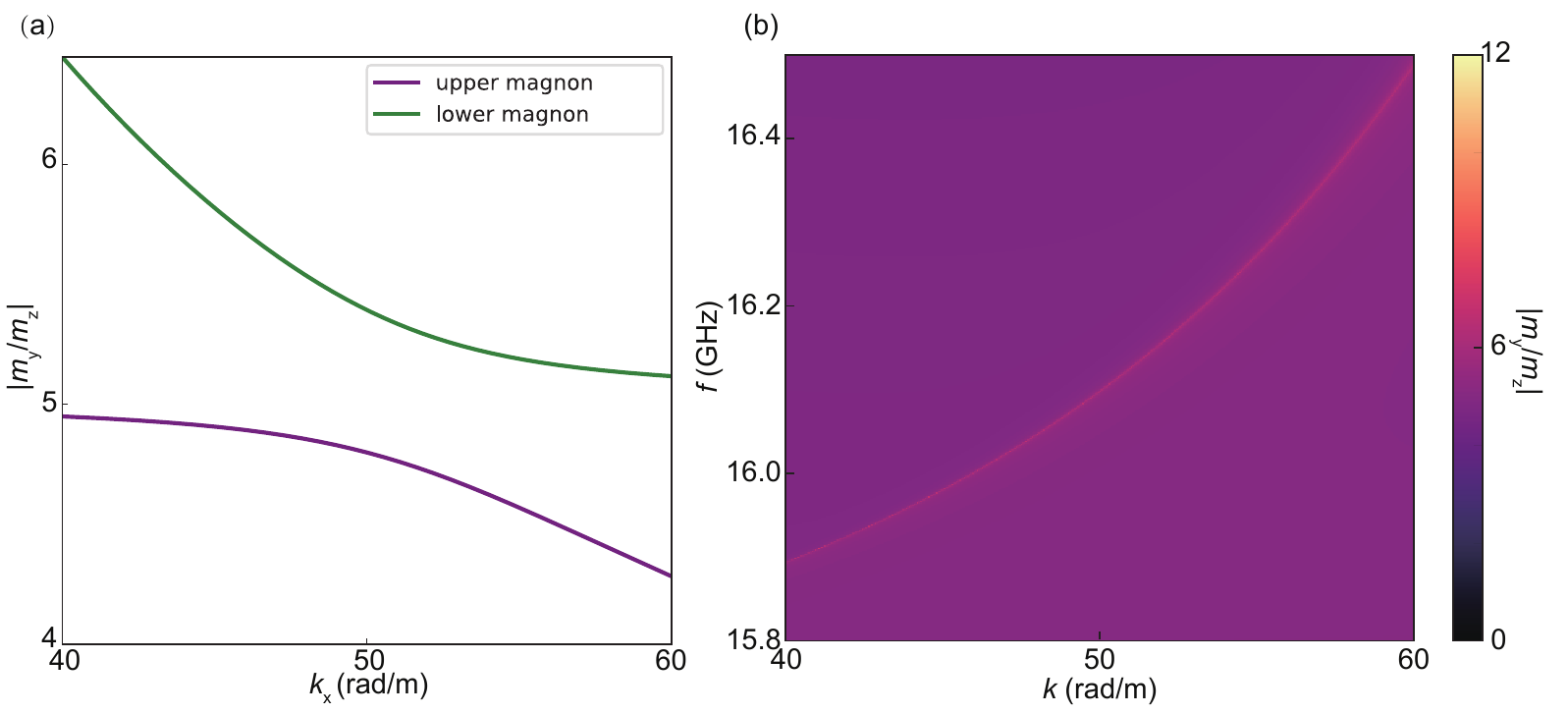}
    \caption{{\bf Wavevector-dependent polarization of bulk and surface magnon-polaritons.} The ratio of the semi-major to semi-minor axis of the precession ellipse of the net magnetic moment $\boldsymbol{m}$, plotted as a function of the wavevector $k$, quantitatively characterizing the polarization of the elliptical precession across different propagation directions.}
    \label{polarization_fig}
\end{figure}
This section focuses on the characters of the net magnetic moment's precession polarization, majorly presented by the ratio $m_y/m_z$.
Fig.~\ref{polarization_fig}(a) show two curves: the upper curve represents the lower magnon, which decreases monotonically from 6.5 to 5, whle the lower curve, corresponding to the upper band, shows a monotonic decrease starting from 5. Similar to the dispersion curves, an anticrossing phenomenon is observable in this figure. The physical origin of this anticrossing lies in the interaction between magnons and photons. In hematite, the anisotropic magnon-photon coupling, described by LLM equations, plays a crucial role. At the anticrossing point, strong magnon-photon hybridization occurs, leading to a non-negligible out-of-plane component ($\boldsymbol{b}_z$) of the photon’s oscillating magnetic field, which affects the magnetic moment precession, altering the ratio $m_y/m_z$.
Specifically, in the anticrossing region, the $m_y/m_z$ is approximately 6, which corresponds to the fact that the band gap in the$ k_y $ direction is six times that in the $ k_x(k_z)$ direction. Increasing $k_x$ shifts the mode toward magnon dominance, the DMI stabilizes this regime and, as a result, the ratio decreases. 
Rather than suppressing $m_z$, it shifts the relative amplitudes of $m_y$ and $m_z$, so the $m_y/m_z$ ratio decreases and deviates from 6. 
This behavior validates the conclusion that anisotropic coupling strength stems from the precession geometry of $\boldsymbol{m}$.

Fig.~\ref{polarization_fig}(b) is a heatmap of $m_y/m_z$ in the frequency-wavevector dispersion plane, covering the energy range of the bulk magnon-photon gap and adjacent bulk modes. 
Two distinct regions are clearly distinguishable in the figure: the first is the bulk mode region, encompassing areas with lower frequencies (lower bulk band) and higher frequencies (upper bulk band), where $m_y/m_z$ exhibits uniformly low values (2–5) with minimal variation across $k_x$. The second is the bulk band gap region, which contains a narrow bright yellow strip. The trajectory of this strip in the plane perfectly matches the dispersion relation of the topological surface magnon-polariton, derived from Eq. 5 in the main text, and within the strip, stabilizes at approximately 10. The significant difference in the ratio between surface and bulk modes arises from the topological origin and interface-bound nature of surface magnon-polariton: the vacuum environment lacks a magnetic medium to support $m_z$-coupled oscillations, leading to the suppression of $m_z$. 
In contrast, $m_y$  can couple to the evanescent wave of photons at the interface—sustained by the dielectric response of hematite, resulting in the enhancement of $m_y$ and the final ratio of $m_y/m_z\sim10$.

\section{Details for surface-mode calculations}\label{surface_A}
In this Appendix, we exhibit the details for the calculations of surface magnon-polariton mode.
Our method follows explicitly of Ref.~\cite{macedo2019engineering}.

First, we give the magnetic susceptibility $\chi(\omega)$.
In Eq.~\ref{LL}, we have $\boldsymbol{m}_{1,2}(\omega)=\frac{1}{i\omega}{\boldsymbol{m}}_{1,2}(\omega)\times \boldsymbol{H}_{\text{eff}}^{1,2}$, which can be further simplified as
\begin{equation}
    \begin{pmatrix}
        m_{1,2}^{y}\\
        m_{1,2}^{z}
    \end{pmatrix}
    =  \begin{pmatrix}
        \chi_{1,2}^{yy} & \chi_{1,2}^{yz}\\
        \chi_{1,2}^{zy} & \chi_{1,2}^{zz}
    \end{pmatrix}
    \cdot
    \begin{pmatrix}
        h^y \\ h^z
    \end{pmatrix},
\end{equation}
where we use the intrinsic frame of references for $\boldsymbol{m}_{1,2}$.
We then transform the susceptibility to the $\boldsymbol{m},\boldsymbol{n}$ basis reads
\begin{equation}
    \left\{
    \begin{aligned}
        m^y&= \left(m_1^y -m_2^y\right)\cos{\theta},\\
        m^z&=m_1^z+m_2^z,
    \end{aligned}
    \right.
\end{equation}
where we can derive the total magnetic susceptibility $\boldsymbol{m}=\chi\boldsymbol{h}$.
Thus, we have the total permeability reads
\begin{equation}
    \mu(\omega)= 
    \begin{pmatrix}
        \mu_{11} & 0 & 0\\
        0 & \mu_{22} & \mu_{23} \\
        0 & \mu_{32} & \mu_{33}
    \end{pmatrix}=\text{diag}\left(\mu_{11}, I_{2}+\mu_0 \chi\right),
\end{equation}
which is useful in the derivation of the dispersion.
Here we note that $\mu_{11}$ is irrelavent to the calculation of dynamic modes in our work, and $I_2$ denotes unit matrix.
Then, we give the methodology to calculate the effective density of states in our model by the semiclassical approach.
For the surface mode of magnon-polaritons in insulators, we have the Ansatz of magnetic scalar vector
\begin{equation}
    \begin{aligned}
    \psi_{\text{in}}&=\psi_{0,in} e^{qz} e^{i(kx-\omega t)},\\
    \psi_{out}&=\psi_{0,out} e^{-\beta z} e^{i(kx-\omega t)}
    \end{aligned}
\end{equation}
where we assume the propagation direction is $x$, and we are considering the upper surface of the slab.
$\nabla \cdot \boldsymbol{B}=0$ in the vacuum gives the $\beta = \sqrt{k^2-\frac{\omega^2}{c^2}}$ in the Ansatz.
At the surface, we have two continuity conditions that the electric displacement vector normal to the surface, and the magnetic field parallel to the surface are continuous. 
Substituting the Ansatz to the Maxwell’s equation $\nabla \times \boldsymbol{E} = -\partial\boldsymbol{B}/\partial t, \nabla\times\boldsymbol{H}=-\partial\boldsymbol{D}/\partial t$, we then obtain~\cite{macedo2019engineering}
\begin{equation}
    \frac{\omega^{2}}{c^{2} \beta} \cdot \frac{-q \mu_{33}(\omega)+i k \mu_{23}(\omega)}{\mu_{22}(\omega) \mu_{33}(\omega)-\mu_{23}(\omega) \mu_{32}(\omega)}=1,
\end{equation}
which can be further simplified as Eq.~\ref{SurfDis}, and the left-hand side is the effective density of states of the surface magnon-polaritons. 
eDOS goes to 1 represents the appearance of coherent mode~\cite{macedo2019engineering}. 
If we substitute the plane-wave Ansatz that $\psi_{\text{in}}=\psi_{0,in} e^{qz} e^{i(kx-\omega t)}$ , with the Maxwell’s equation inside the bulk, we obtain
\begin{equation}
    \frac{1}{\epsilon \omega^{2}} \cdot \frac{k^{2} \mu_{33}(\omega)}{\mu_{22}(\omega) \mu_{33}(\omega)-\mu_{23}(\omega) \mu_{32}(\omega)}=1,
\end{equation}
whose left-hand side is also a function of $\omega$ and $k$. 
If we plot the left-hand side, the maxima which hit the quantity $1$ gives the solution to the bulk magnon-photon mode.
\begin{figure*}
    \includegraphics[width=0.88\textwidth]{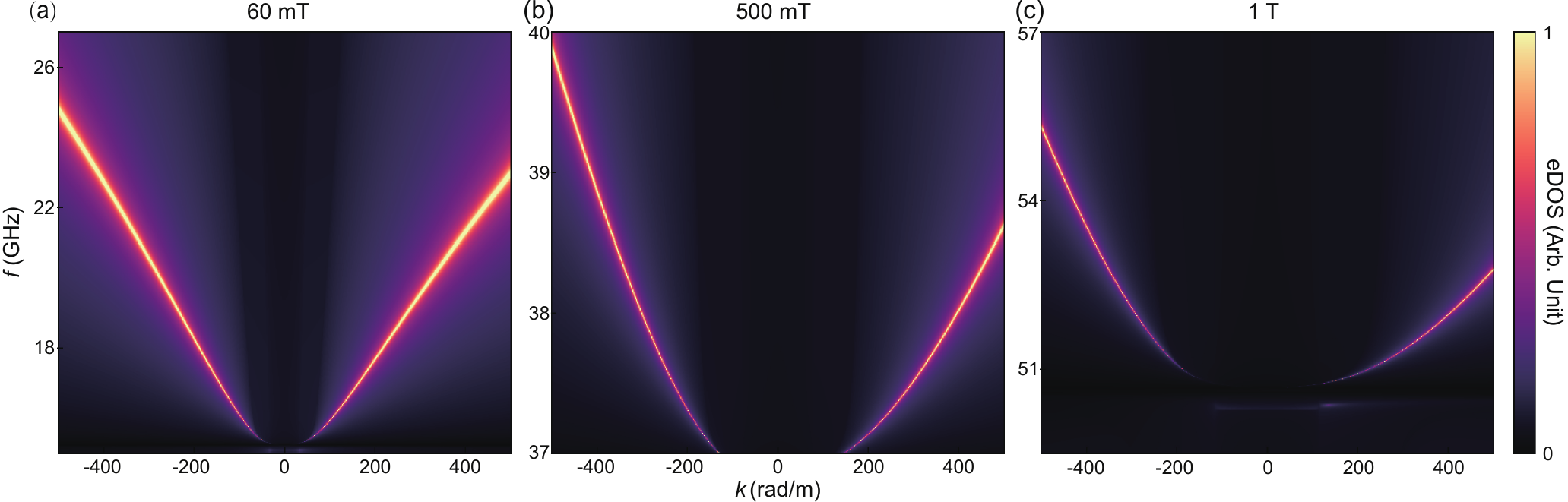}
    \caption{{\bf Magnetic field tunability of nonreciprocal surface magnon-polaritons.} Effective density of states (eDOS) of surface magnon-polaritons under different external magnetic field configurations: (a) for $50~\text{mT}$, (b) for $500~\text{mT}$, and (a) for $1~\text{T}$, demonstrating the nonreciprocal dispersion and its sensitivity to external magnetic fields.}
    \label{NR_fig}
\end{figure*}
\section{Magnetic field tunability of nonreciprocal surface magnon-polaritons}\label{NR_A}

This section supplements the modulation of topological surface states by external magnetic fields, focusing on the magnetic field response of nonreciprocal surface magnon-polaritons. 
Fig.~\ref{NR_fig} shows the effective density of states (eDOS) of surface magnon-polaritons under three distinct external magnetic field strengths $(H_0=60~\text{mT}, H_0=500~\text{mT}, H_0=1~\text{T})$, consistent with the experimental field range specified in the main text. 
In these eDOS maps, peak intensity characterizes mode coherence, while asymmetry relative to quantifies the degree of nonreciprocity (a core feature of the topological surface modes discussed in the main text).
The two-sublattice system is fundamentally characterized by the DMI, which tilts the magnetization of both sublattices. 
This tilted state imparts intrinsic nonreciprocal behavior to the system. 
Notably, this nonreciprocal characteristic exhibits high sensitivity to variations in the external magnetic field $H_0$. 
Direct observations from Fig.~\ref{NR_fig} (corresponding to the three aforementioned values) reveal a clear trend: as $H_0$ increases, the asymmetry of the eDOS peaks becomes increasingly pronounced. 
Specifically, the eDOS peaks at $H_0=60\text~{mT}$ show relatively weak directional asymmetry; at $H_0$=500 mT, the peaks exhibit a more significant bias toward one propagation direction; and at $H_0=1\text~{T}$, this asymmetric feature is further enhanced, with the eDOS peak asymmetry reaching its maximum within the measured field range.

The physical origin of this trend lies in the modulation of sublattice magnetization tilted by DMI and external field.
As $H_0$ increases, it combines with the intrinsic $H_{\text{DM}}$, further amplifying the asymmetry of the magnetization distribution. 
This more pronounced magnetization asymmetry, in turn, enhances the directional difference in the coupling between surface magnon-polaritons and the underlying magnetic order: the mode propagates more strongly in one direction (corresponding to higher frequency) and more weakly in the opposite direction (corresponding to lower frequency). 
This mechanism directly explains the increasingly asymmetric eDOS peaks in Fig.~\ref{NR_fig} and provides experimental support for the main text’s conclusion that nonreciprocal topological surface modes can be tuned by external magnetic fields.

\begin{figure}
    \includegraphics[width=0.48\textwidth]{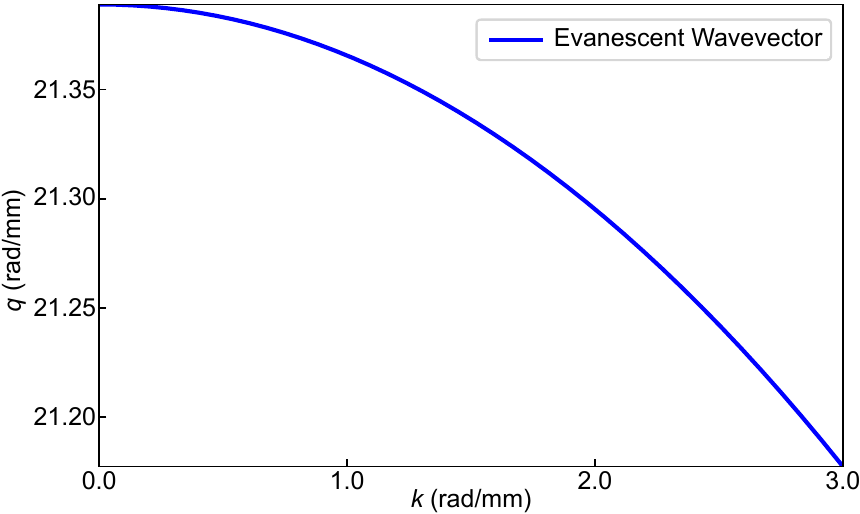}
    \caption{{\bf Matching the evanescent wavevector $q$.} The evanescent wavevector $q$ as a function of the in-plane wavevector $k_x$, characterizing the evanescent wavevector should be matched by the microwave antenna for experimental realizations of surface magnon polaritons.}
    \label{matching_fig}
\end{figure}
\section{Matching between propagating and evanescent wavevector}\label{match_A}

The excitation of the topological surface magnon-polariton requires phase matching to its in-plane wavevector k and field profile matching to its evanescent decay in the vacuum, characterized by the wavevector $q$. 
Fig.~\ref{matching_fig} plots the calculated evanescent wavevector $q$ as a function of $\boldsymbol{k}$ for the surface magnon-polariton mode. 
The observed monotonic decrease of $q$ with increasing $k$ is a direct consequence of the surface magnon-polariton dispersion relation. 
As the in-plane wavevector $k$ increases, the frequency of the surface magnon-polariton mode approaches the light line of the dielectric medium. 
Near this limit, the electromagnetic fields of the mode become less confined to the material interface and extend further into the vacuum, which corresponds to a diminishing spatial decay rate, or a smaller $q$. 
This relationship is critical for experimental design. 
A smaller $q$ at larger $k$ implies a longer evanescent tail in the vacuum, significantly facilitating the coupling of external excitation sources, such as a prism or an antenna, to the surface mode. 
Therefore, the operational regime at larger in-plane wavevectors is advantageous for the efficient excitation of the topological surface magnon-polariton, as the required spatial matching of the evanescent field is more readily achievable.
\bibliography{references}

\begin{thebibliography}{70}
\bibitem [{not()}]{note1}%
  \BibitemOpen
  \href@noop {} {}\bibinfo {note} {Any value of $\boldsymbol{k}$ range for integral (for example, -100~100 rad/m) that include all $\boldsymbol{k}$ points with nonzero Berry curvature gives the same result here.}\BibitemShut {Stop}%
\end{thebibliography}

\end{document}